\documentclass[aps,prl,twocolumn,superscriptaddress,showpacs]{revtex4}
\usepackage{graphicx}
\begin{document}
\newcommand{\bea}{\begin{array}}
\newcommand{\ena}{\end{array}}

\title{Non-ergodicity and localization of invariant measure for two colliding masses}

\author{Jiao Wang}
\affiliation{Department of Physics and Institute of Theoretical Physics
and Astrophysics, Xiamen University, Xiamen 361005, Fujian, China}
\author{Giulio Casati}
\affiliation{CNISM and Center for Nonlinear and Complex Systems,
Universit\`a degli Studi dell'Insubria, via Valleggio 11, 22100 Como, Italy}
\affiliation{Istituto Nazionale di Fisica Nucleare, Sezione di Milano, via
Celoria 16, 20133 Milano, Italy}
\author{Toma\v z Prosen}
\affiliation{Physics department, Faculty of Mathematics and Physics,
University of Ljubljana, Ljubljana, Slovenia}
\date{\today}

\begin{abstract}
We show evidence, based on extensive and carefully performed numerical experiments,
that the system of two elastic hard-point masses in one-dimension is not ergodic
for a generic mass ratio and consequently does not follow the principle of energy
equipartition. This system is equivalent to a right triangular billiard. Remarkably,
following the time-dependent probability distribution in a suitably chosen velocity
direction space, we find evidence of exponential localization of invariant measure.
For non-generic mass ratios which correspond to billiard angles which are rational,
or weak irrational multiples of $\pi$, the system is ergodic, in consistence with
existing rigorous results.
\end{abstract}

\maketitle

{\it Introduction.--} Soon after the Sinai's proof~\cite{Sinai70} of ergodicity
and mixing in the two-dimensional hard disc gas, the question has been raised if
ergodicity and mixing may not also occur in an even simpler one-dimensional, unequal
mass, hard point gas. In particular, as pointed out by Lebowitz long ago~\cite{Leb71},
even the simplest unequal mass case having only two moving point particles is not
decided. Since those pioneering times, the analysis of simple non-trivial models
has allowed a tremendous progress in our understanding of the properties of nonlinear
dynamical systems. In particular, the above problem is of special interest since the
local dynamical instability is only linear and therefore it is worthwhile to inquire
to what extent statistical properties are present in such systems. This problem is
also relevant for the understanding of the properties of the diffusion and relaxation
process in quantum mechanics. Indeed, unlike the exponentially unstable classical
chaotic motion, in the quantum case deviations in the initial conditions propagate
only linearly in time and therefore the quantum diffusion and relaxation process
takes place in the absence of exponential instability.

The dynamics of the one-dimensional unequal mass, hard point gas with reflecting
boundary conditions can be reduced to a simple map. Indeed let $m_1,m_2$ be the
masses of the two particles, $x_1,x_2$ and $v_1,v_2$ their positions and velocities
respectively. After the collision, the new velocities $v'_1,v'_2$ are given by
\begin{eqnarray}
\left(\bea{c} {v'_1}\\v'_{2}\ena\right)=\left(\bea{lc}
\cos\alpha &1- \cos\alpha\\ 1+\cos\alpha & -\cos\alpha\ena
\right)\left(\bea{c} {v_1}\\v_{2}
\ena\right),
\end{eqnarray}
where
\begin{eqnarray}
\cos \alpha=\frac{m_1-m_2}{m_1+m_2}.
\end{eqnarray}
To simplify the dynamical description it is convenient to introduce the rescaling
$y_i=\sqrt{m_i}x_i$,  $w_i=\sqrt{m_i}v_i,(i=1,2)$. Then by letting $\tan \theta=
w_2/w_1$, Eq.~(1) can be written as $\tan(\theta') = \tan(\alpha-\theta)$. The
collisions with the left and right boundary only imply inversion of a particle's
velocity and are described by $\theta'=\pi-\theta$ or $\theta'=-\theta$, respectively.
It is interesting to notice that our model is equivalent to a point particle with
coordinate $(y_1,y_2)$ and velocity $(w_1,w_2)$ moving inside a right triangular
billiard with one acute angle given by $\alpha/2$.

If we now denote by $C$ the particle collision, $L$ and $R$ the collision with the
left and right wall respectively, it is easy to see that $CR$ merely rotates the
velocity vector $w =(w_1,w_2)$  by the angle $\alpha$ while $RC$ performs the inverse
rotation. Taking into account that $L=-R$, the overall dynamics in the velocity space
can be described as a rotation by an angle $\pm\alpha$ where the sequence of
left/right rotations is determined by the given orbit. In other words, for a given
trajectory starting with the initial value $\theta_0$, the final value $\theta$, after
a finite sequence,  can be written as $\varphi+K \alpha$ with $K$ being an integer and
$\varphi$ being one of the following four constant values: $\theta_0$, $-\theta_0$,
$\pi-\theta_0$, and $\theta_0-\pi$ where the four different values correspond to the
four different signs in the velocity pair. In terms of the integer $K$, the dynamics
merely reduces to the map $K'=-K+1$ (particles collision) and $K'=-K$ (boundary
collision).

This model has been extensively studied both analytically and numerically and in
spite of its seemingly simplicity, no definite conclusion has been reached so far
concerning its dynamical properties even though the prevailing opinion is that, for
$\alpha$ being irrational multiple of $\pi$, the system should be ergodic and weakly
mixing. Indeed if $\alpha/\pi$ is rational, i.e. $\alpha= 2\pi p/q$ with coprime
integers $p$ and $q$, then there are precisely $4q$ distinct velocity pairs allowed
which means that $K$ can take only $q$ different integer values. On the other hand,
when $\alpha/\pi$ is irrational, the allowed $\theta$ values become uniformly dense
in the interval $[0,2\pi]$ and it is at least possible for the system to be ergodic
in the velocity space. This latter possibility has actually been suggested long ago
on the basis of the first numerical investigations of the model~\cite{GCJF76} even
though the authors themselves pointed out the surprisingly large total number of
collisions required to observe all the allowed $q$ values of $K$. Subsequent numerical
investigations~\cite{ACG97} confirmed this original conclusion. In a more recent paper
~\cite{CP99mix}, the case of a generic triangular billiard with all angles irrational
has been considered. (The dynamics of a point particle in this billiard is equivalent
to the motion of three hard-core point particles on a ring~\cite{CP99mix,Glashow}.)
Here, while empirical evidence strongly demonstrates that irrational triangular
billiards are mixing, it is pointed out that no definite conclusions can be made
concerning right triangular billiards, a problem which still remains open.

On the analytical side, the problem has attracted a lot of interest in the mathematical
community \cite{Gutkin86}. For a review we refer to Ref.~\cite{Gutkin96}. In particular,
in Ref.~\cite{SK86math} the rational case is considered for which the orbits with initial
$\theta_0$ lie on an invariant surface $M_{{\theta}_0}$ determined by the finite number
of velocities. A theorem in \cite{SK86math} states that for almost every $\theta_0$ the
flow  $f_t|M_{{\theta}_0}$ is uniquely ergodic. This result indicates that in the
irrational case an orbit densely covers the energy surface. However, it does not allow
to conclude that the system is ergodic. Indeed, it is well known that the quasi-ergodic
hypothesis, namely, that an orbit comes arbitrarily close to every point on the energy
surface, is not sufficient to guarantee ergodicity, namely, that time averages are equal
to phase space averages.  A more recent, interesting result in Ref.~\cite{vorob96}
proves that the system is ergodic for irrational values of  $\alpha/\pi$ which are very
quickly approximated by sequences of rational numbers. The interesting problem remains
open of what happens for a typical (generic) irrational value of $\alpha/\pi$.

It is the purpose of the present paper to examine this question with the help of a
highly accurate numerical analysis unaccessible previously. The result turns out to
be very surprising: numerical evidence strongly suggests that for a typical irrational
value of $\alpha/\pi$, the system is non-ergodic with typical invariant measures being
exponentially localized in the integer direction index $K$.

Without loss of generality, we fix the length of the box to be unity, the mass of
the first particle to be unity, $m_1=1$, and the total energy of the system to be
$E_1+E_2=1/2$.

{\it The `strong' irrational case.--} We consider a mass ratio corresponding to the
most irrational case $\alpha=\alpha_g$ with $\alpha_g/\pi=(\sqrt{5}-1)/2$. In order
to investigate its dynamical properties we consider a sequence of principal convergents
$\alpha/\pi=M/N$ in the continued fraction representation; namely,$M/N$=  $[0,1,1,1,1,
\cdots]$ = $1/2$, $2/3$, $3/5$, $5/8$, $8/13$, $\ldots$, $f_i/f_{i+1}$, $\ldots$,
where $f_{i+1}=f_i+f_{i-1}$. The idea is that the dynamical behavior of each principal
convergent might allow perhaps to draw some conclusions for the limiting $N\to\infty$,
golden mean case $\alpha=\alpha_g$.

A necessary condition for ergodicity is certainly an equipartition of energy in
velocity space. Therefore, for a given principal covergent, it is natural to evaluate
the time required for an orbit to equally populate, in time average, all the allowed
($2N$ for odd $M$ or $N$ for even $M$) values of $K$. It turns out that this time is
exceedingly long and already for a very low approximant it goes beyond accessible
computing resources.

%%%%%%%%%%%%%%%%%%%%%%%%%%%%%%%%%%%%%%%%%%%%%%%%%%%%%%%%%%%%%%%%%%%%%%%%%%%%%%%%%%%%%%%%%%%% Fig 1
\begin{figure}[!t]
\hspace{-.8cm}
\includegraphics[width=1.\columnwidth,clip]{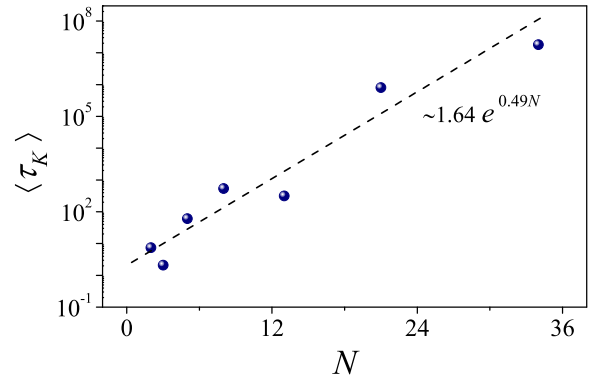}\vspace{-0.5cm}
\caption{(Color online) The ensemble averaged $\tau_K$ for eight low order rational
approximants of $\alpha/\pi=(\sqrt{5}-1)/2$. The ensemble consists of $10^4$
trajectories with fixed $\theta_0=e\pi$ but different initial particle positions
randomly assigned. The error of each data point (not shown) is smaller than the
symbol. The numerical results suggest that $\langle\tau_K \rangle$ has an exponential
dependence on $N$. The dashed line indicates the result of the best linear fitting:
$\langle\tau_K \rangle \sim e^{0.49N}$.}
\end{figure}
%%%%%%%%%%%%%%%%%%%%%%%%%%%%%%%%%%%%%%%%%%%%%%%%%%%%%%%%%%%%%%%%%%%%%%%%%%%%%%%%%%%%%%%%%%%% Fig 1

We then inquire about a shorter, relevant time scale, denoted by $\tau_K$ and defined
as the shortest time for a trajectory to visit all the allowed values of $K$. Since
$\tau_K$ wildly fluctuates from a trajectory to another, we compute the average $\langle
\tau_K \rangle$ over an ensemble of trajectories with different initial particle positions.
(The initial velocities are fixed.) The results, presented in Fig. 1, show that $\langle
\tau_K \rangle$ increases exponentially with $N$ thus implying that for the golden mean
case this time diverges to infinity astonishingly fast. One may argue that the question
of ergodicity for such a system may be an abstract issue, physically irrelevant: even
for a low convergent $M/N$ with $N=500$, if one assumes that there are $10^{10}$ collisions
per second, the time required to visit all the allowed values of velocities would be much
larger than the age of the universe. Nevertheless, abstract models are known to play an
important role in the theory of dynamical systems. We introduce therefore a second, much
shorter time scale $\tau^*$ which will allow to approach much closer to the golden mean
via the sequence of principal convergents.

%%%%%%%%%%%%%%%%%%%%%%%%%%%%%%%%%%%%%%%%%%%%%%%%%%%%%%%%%%%%%%%%%%%%%%%%%%%%%%%%%%%%%%%%%%%% Fig 2
\begin{figure}[!t]
\vspace{-0.1cm}\hspace{-.8cm}
\includegraphics[width=1.\columnwidth,clip]{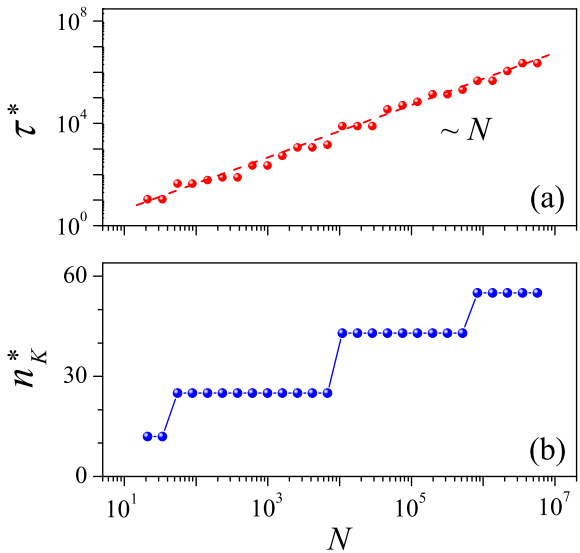}\vspace{-0.5cm}
\caption{(Color online) (a) The dependence of $\tau^*$ on $N$, in the continued fraction
rational approximation $M/N$ of $\alpha/\pi=(\sqrt{5}-1)/2$, for a given, fixed, initial
condition of $\theta_0=e\pi$ and randomly assigned particle positions. The dashed line
$\sim N$ is plotted for reference. For each rational approximation, the number  $n^*_K$
of different $K$ values visited by the trajectory up to time $\tau^*$ is shown in (b).}
\end{figure}
%%%%%%%%%%%%%%%%%%%%%%%%%%%%%%%%%%%%%%%%%%%%%%%%%%%%%%%%%%%%%%%%%%%%%%%%%%%%%%%%%%%%%%%%%%%% Fig 2

Let us consider two subsequent principal convergents $M/N$  and $M'/N'$ (with $N'>N$);
the two trajectories obtained by evolving the same initial condition with $\alpha/\pi=M/N$
and $M'/N'$ respectively, would follow the same symbolic collision sequence (of letters
$C$, $L$, $R$) up to some time $\tau^*(N,N')$. This means that up to time  $\tau^*(N,N')$,
the sequence of collisions will be the same as for the limiting golden mean case. In
Fig.~2(a) we plot $\tau^*(N,N')$ for a given fixed initial condition where $\theta_0/\pi=e$
and the particle positions are assigned randomly. Despite fluctuations, it can be seen that,
fairly accurately, $\tau^*$ increases linearly with $N$. In Fig.~2(b), the number $n^*_K$ of
different $K$ values visited by the trajectory up to time $\tau^*$, is shown. Surprisingly,
we notice that while $\tau^*$ increases linearly with $N$, the number $n^*_K$ of actually
visited $K$ values increases very slowly with $N$. Even though it is difficult to determine
with sufficient accuracy the dependence on $N$ due to the presence of sudden jumps and long
plateaus, this dependence seems to be logarithmic $n^*_k\sim\ln(N)$. Therefore, for any
approximant $M/N$ of the golden mean case, the system visits only a small fraction of the
allowed values of direction $K$ up to time $\tau^*\sim N$.

%%%%%%%%%%%%%%%%%%%%%%%%%%%%%%%%%%%%%%%%%%%%%%%%%%%%%%%%%%%%%%%%%%%%%%%%%%%%%%%%%%%%%%%%%%%% Fig 3
\begin{figure}[!t]
\hspace{-.8cm}
\includegraphics[width=1.\columnwidth,clip]{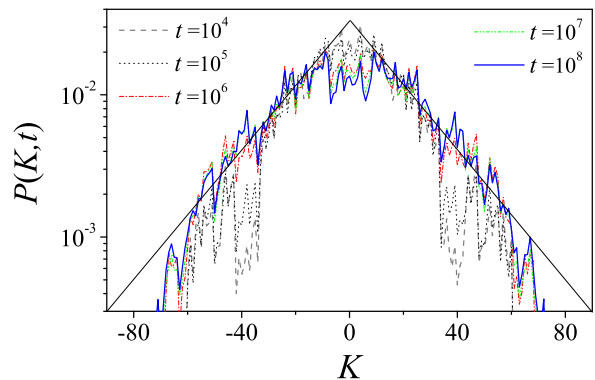}\vspace{-0.5cm}
\caption{(Color online) The probability density distribution function $P(K,t)$ computed
at various times $t$ for the strong irrational case with $\alpha/\pi$ being the golden
mean. Starting from the lower tail, the curves are for $t=10^4$ (dashed), $10^5$
(dotted), $10^6$ (dash-dotted), $10^7$ (dash-dot-dotted), and $10^8$ (solid),
respectively. The thin black lines are drawn for reference. $\theta_0=\pi/\sqrt{2}$,
$\delta t=t/10$, and an average ensemble of 5200 trajectories is taken for evaluating
$P(K,t)$ (see the text). }
\end{figure}
%%%%%%%%%%%%%%%%%%%%%%%%%%%%%%%%%%%%%%%%%%%%%%%%%%%%%%%%%%%%%%%%%%%%%%%%%%%%%%%%%%%%%%%%%%%% Fig 3

We now turn to the direct consideration of the irrational case $\alpha=\alpha_g$. For
this case we compute the probability density distribution function $P(K,t)$ obtained
by evolving an ensemble of trajectories with the same initial particle velocities but
with random initial positions in the interval [0,1]. In order to suppress fluctuations
we make an additional, partial time average of $P(K,t)$ in the interval $[t,t+\delta t]$.
The results are shown in Fig.~3, where we plot $P(K,t)$ at different times $t$, from
$10^4$ to $10^8$. The main outcome is that the distribution $P(K,t)$ remains very narrow
(of width $\sim100$) even at very large times. As time increases the tails rise up and
one can observe an overall tendency to an exponential distribution. In Fig.~4 we plot
the variance of $P(K,t)$. It is seen that the variance appears to saturate even though
the convergence is very slow probably due to the slow convergence of the tails to the
exponential distribution seen in Fig.~3. The scenario which seems to emerge from
Figs.~2-4 is that the system gradually ``sees" the successive principal convergents of
the golden mean limiting case. Correspondingly, the distribution slowly converges to
an exponential shape.

%%%%%%%%%%%%%%%%%%%%%%%%%%%%%%%%%%%%%%%%%%%%%%%%%%%%%%%%%%%%%%%%%%%%%%%%%%%%%%%%%%%%%%%%%%%% Fig 4
\begin{figure}[!b]
\hspace{-.8cm}
\includegraphics[width=1.\columnwidth,clip]{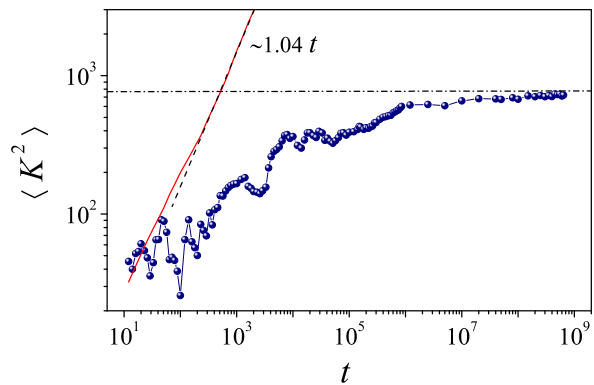}\vspace{-0.5cm}
\caption{(Color online) The time dependence of the variance of $P(K,t)$ for the data
of Fig.~3 (bullets). The full (red) curve is the counterpart of the modified model
with randomized positions. The horizontal dash-dotted line is the expected saturation
value corresponding to the exponential distribution in Fig.~3. The dashed line corresponds
to $\langle K^2\rangle = 1.04 t$.
}
\end{figure}
%%%%%%%%%%%%%%%%%%%%%%%%%%%%%%%%%%%%%%%%%%%%%%%%%%%%%%%%%%%%%%%%%%%%%%%%%%%%%%%%%%%%%%%%%%%% Fig 4

One should be very careful in drawing definite conclusions from numerical computations.
However the following results, namely, (i) the fraction of different $K$ values visited
by an orbit up to time $\tau^*$ is negligibly small, of order $\ln(N)/N$, while $\tau^*
\sim N \rightarrow\infty$ in the golden mean limit, and (ii) the probability distribution
$P(K,t)$ has an exponential shape in the limit $t\to\infty$, strongly suggest that the
golden mean case is non-ergodic.

It is interesting to compare the dynamics of our system with the behavior of the
following modified model: after each collision we randomize the particle positions for
all trajectories. In this case it turns out that the variance $\langle K^2\rangle\sim Dt$
with $D\approx 1$ (full line in Fig.~4). The resulting probability distribution $P(K,t)$
is shown in Fig.~5 (full line); here the dash-dotted line is a Gaussian with $D\approx 1$
and it agrees quite well with numerical data. The striking difference with the distribution
of the actual model (dashed line), plotted after the same time, shows the strong freezing
effect induced by dynamical correlations.

%%%%%%%%%%%%%%%%%%%%%%%%%%%%%%%%%%%%%%%%%%%%%%%%%%%%%%%%%%%%%%%%%%%%%%%%%%%%%%%%%%%%%%%%%%%% Fig 5
\begin{figure}[!t]
\hspace{-.8cm}
\includegraphics[width=1.\columnwidth,clip]{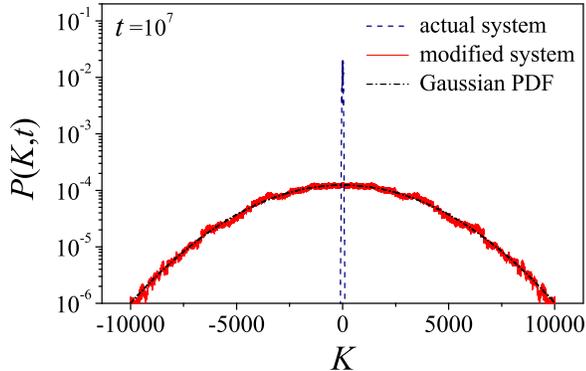}\vspace{-0.5cm}
\caption{(Color online) Probability distribution function $P(K,t)$ for the modified
model with randomized particle positions (full curve) compared with the actual
distribution for the dynamical model (dashed line). The two distributions are plotted
at the same time $t=10^7$. The dash-dotted curve is the Gaussian distribution for
$D=1.04$ and $t=10^7$. }
\end{figure}
%%%%%%%%%%%%%%%%%%%%%%%%%%%%%%%%%%%%%%%%%%%%%%%%%%%%%%%%%%%%%%%%%%%%%%%%%%%%%%%%%%%%%%%%%%%% Fig 5

{\it Weak irrational case.--}
So far we have considered a particular mass ratio given by $\alpha=\alpha_g$. The
natural question is how generic is the dynamical behavior displayed by this case.
To this end we recall an important rigorous result by Vorobets~\cite{vorob96} which
states that the billiard flow in a right triangular with one acute angle $\alpha/2=
\pi(a_5^{-1}+a_{10}^{-1}+...+a_{5n}^{-1}+\ldots)$, where the sequence ${a_n}$ is
given by $a_0 =1$ and $a_{n+1}= 2^{a_n}$ ($n= 0,1,2,....$), is ergodic. Therefore,
ergodicity takes place for an irrational $\alpha/\pi$ which is very close to
rational numbers in the sense that the sequence of rational approximants has an
impressively fast convergence. This result is in nice agreement with our numerical
analysis. Indeed, given two subsequent principal convergents $M_i/N_i, M_{i+1}/N_{i+1}$
of an irrational $\alpha/\pi$, in order to be ergodic it is at least necessary that
the time $\tau^*(N_i,N_{i+1})$ is larger than the time $\tau_K (N_i)$ needed to visit
all the allowed velocities at the $i$-th rational approximant. Since, as we have
seen, $\tau_K$ increases exponentially with $N$ while $\tau^*$ increases only linearly,
it follows that a necessary condition for ergodicity of the limiting irrational $\alpha$
is that $N_{i+1}\sim\exp{(N_i)}$. Therefore our numerical analysis not only is in
agreement with the rigorous result~\cite{vorob96} but also provides a hint for
identifying the set of irrational $\alpha/\pi$ for which the system can be ergodic.
These irrationals are inside the so called set of Liouville numbers~\cite{Li} which
are known to have zero Lebesgue measure.

{\it Discussion.--}
It is interesting to remark that the predicted lack of ergodicity is not related
to the usual picture of divided phase space with islands of stability and region
of chaotic motion. Moreover, in our case, following the result of Ref.~\cite{SK86math},
a typical orbit is expected to densely cover the energy surface. Instead the non-ergodic
behavior derives from the exponential decay of the probability distribution $P(K,t)$
indicating exponential localization of invariant measures. This feature explains why
previous numerical investigations failed to identify non-ergodic behavior. The reason
is not the larger computational power now available; instead it is the dynamical
quantities that were investigated. To make a clear example, by the popular surface
of section method it would be absolutely impossible to provide numerical evidence
of non-ergodic behavior in a system where the orbits are dense on the energy surface.

The behavior of our model is reminiscent of the dynamical localization phenomenon
in the quantum kicked rotator~\cite{GC79} in which the relevant parameter is the
kicking period $\tau$. Indeed it is known that for a typical irrational $\tau/\pi$,
exponential localization takes place while it has been rigorously proven that there
exist irrational values of period $\tau/\pi$ for which there is no localization~\cite{GC84}.
Actually, heuristic arguments~\cite{CC95} lead to the conclusion that such irrationals
belong to the set of Liouville numbers. Quite clearly this analogy should be taken
\emph{cum grano salis }since the two systems belong to completely different domains.
Yet they have in common that a key role is played by the randomness property of the
sequence of digits of the irrational $\alpha/\pi$ or $\tau/\pi$. This may open the
way for a better understanding of the mechanism leading to the suppression of
classical and quantum diffusion. In addition, one is confronted one more time with
the fascinating richness of dynamical behavior of seemingly extremely simple systems.

$Acknowledgements.-$ We are grateful to I. Guarneri and R. Artuso for useful discussions.
J.W. acknowledges the support by the NNSF (Grant No. 11275159) and SRFDP (Grant No.
20100121110021) of China, G.C. acknowledges the support by MIUR-PRIN 2010-2011 and
by Regione Lombardia, and T.P. acknowledges Grant P1-0044 of the Slovenian Research
Agency.


\begin{thebibliography}{99}
\bibitem{Li}       The Liouville numbers are those numbers which are approximated by rationals $M/N$ faster than any power of $N$.
\end{thebibliography}
\end{document}